# The Growing Importance of a Tech Savvy Astronomy and Astrophysics Workforce

Lead: Dara Norman, NOAO, dnorman@noao.edu
Co-Is:
Kelle Cruz, Hunter College, kellecruz@gmail.com, Vandana Desai, Caltech/IPAC-IRSA, desai@ipac.caltech.edu, Britt Lundgren, University of North Carolina, Asheville, blundgre@unca.edu , Eric Bellm, University of Washington, ecbellm@uw.edu, Frossie Economou, LSST, frossie@lsst.org, Arfon Smith, STScI, arfon@stsci.edu, Amanda Bauer, LSST, ABauer@lsst.org, Brian Nord, Fermi Lab, nord@fnal.gov, Chad Schafer, Carnegie Mellon University, cschafer@cmu.edu, Gautham Narayan, STScI, gnarayan@stsci.edu, Ting Li, Fermi Lab, tingli@fnal.gov, Erik Tollerud Space Telescope Science Institute, etollerud@stsci.edu, Brigitta Sipőcz, DIRAC Institute, University of Washington bsipocz@gmail.com, Heloise Stevance University of Auckland & University of Sheffield, hfstevance@gmail.com, Timothy Pickering, MMT Observatory, te.pickering@gmail.com, Manodeep Sinha, Swinburne University of Technology and ARC Centre of Excellence for All Sky Astrophysics in 3 Dimensions (ASTRO 3D), msinha@swin.edu.au Joseph Harrington, University of Central Florida, jh@physics.ucf.edu, Jeyhan Kartaltepe, Rochester Institute of technology jeyhan@astro.rit.edu, Dany Vohl, ASTRON, Netherlands Institute for Radio Astronomy, vohl@astron.nl, Adrian Price-Whelan, Center for Computational Astrophysics, Flatiron Institute, adrianmpw@gmail.com, Brian Cherinka, Johns Hopkins University, havok2063@gmail.com, Chi-kwan Chan, Steward Observatory, University of Arizona, chanc@email.arizona.edu, Benjamin Weiner, MMT Observatory, University of Arizona, bjw@mmto.org, Maryam Modjaz, NYU, mmodjaz@nyu,edu, Federica Bianco, University of Delaware, fbianco@udel.edu, Wolfgang Kerzendorf, Michigan State University, wkerzendorf@gmail.com, Iva Laginja, Space Telescope Science Institute, ilaginja@stsci.edu, Chuanfei Dong, Princeton University, dcfy@princeton.edu

Type of Activity: Infrastructure Activity, State of the Profession, Other: Workforce

1. **Executive Summary and Recommendations:**
   **Key Issue and Overview of Impact on the Field**

Fundamental coding and software development skills are increasingly necessary for success in nearly every aspect of astronomical and astrophysical research as large surveys and high resolution simulations become the norm. However, professional training in these skills is inaccessible or impractical for many members of our community. Students and professionals alike have been expected to acquire these skills on their own, apart from formal classroom curriculum or on-the-job training. Despite the recognized importance of these skills, there is little opportunity to develop them - even for interested researchers.

To ensure a workforce capable of taking advantage of the computational resources and the large volumes of data coming in the next decade, we must identify and support ways to make software development training widely accessible to community members, regardless of affiliation or career level. **To develop and sustain a technology capable astronomical and astrophysical workforce, we recommend that agencies make funding and other resources available in order to encourage, support and, in some cases, require progress on necessary training, infrastructure and policies.** In this white paper, we focus on recommendations for how funding agencies can lead in the promotion of activities to support the astronomy and astrophysical workforce in the 2020s.

**1.1 Recommendations for Training and Training Materials Support**

1. Fund more and large-scale programs that cultivate the next generation of researchers versed in astronomy, astrophysics and data science, similar to small and oversubscribed, grass-roots programs.
2. Fund data and computational centers to produce modular software training resources for the community. These resources should be designed to be used by individuals, integrated into formal classes, and used as part of professional development training.

**1.2 Recommendations to Support Infrastructure for Training**

3. Provide funding to host and curate educational materials in a long-term, stable, scalable place. A centralized location will provide stability and improve the discoverability of the material.

4. Provide incentives to launch opportunities to harness innovative partnerships between data centers, universities, and industry. For example, support for sabbatical programs at the data centers where teaching faculty can learn skills, and develop educational materials for themselves and wider community use.
5. Encourage and incentivize individuals, departments, and professional societies to build educational programs that incorporate software training skills into their existing courses and programs.

 **1.3 Recommendations to Encourage Policies around Career Development and Inclusion**

6. Promote and resource big data science training activities and professional development for non-students (i.e., post-docs and career professionals) as part of the science mission and deliverables of federally-funded data centers.
7. Require advisory board representation of federally-funded science and data centers to include representatives from small and under-resourced institutions to provide a broad and clear picture of community needs.

**2. The Need for Tech-Savvy Astronomers**

**In the era of large surveys, experiments, and datasets, we will only reach our scientific goals if we train and retain highly capable scientists who are also engaged with technological advances in computing.** With the goal of advancing scientific discovery through the collection and analysis of increasing amounts of data, we must commit and dedicate resources to building both the skills and competencies of this workforce. We define the workforce as those who build scientific insight from the data, as well as those who build the infrastructure to collect and prepare the data for analysis - one cannot happen without the other. The areas and skill sets in which our teams need training include software carpentry, algorithms, statistics[1] and the use of software tools and user services[2]. Support staff will also need to be versed in software engineering effective practices, as well as, data management and access. In addition, all members of the workforce, including those who provide infrastructure support, will require professional development opportunities and well defined pathways for career advancement that allow them to remain as productive and respected members of the field.

---

[1] For additional details see ASTRO2020 science white paper, ``The Next Decade of Astroinformatics and Astrostatistics'', Aneta Siemiginowska, et al..
[2] For an example see ASTRO2020 science white paper,"Science Platforms for Resolved Stellar Populations in the Next Decade", K.Olsen, et al.

Below we discuss the activities necessary to build, support, and advance the scientific workforce that will take the petabytes of data collected to scientific discoveries over the next decade. In particular, we next discuss the current demographics of the data science support mission which exemplifies the scope of training required to build this workforce. We focus on training for researchers who are more accurately described as "users", as well as, the career paths and training needs of staff who build and support infrastructure. We also discuss modern challenges for these career paths, and ways to address them.

**3. Current Demographics - who is this workforce and where are they**

Data support roles are spread throughout the astronomy and astrophysics (hereafter, ``astronomy") scientific community, and encompass people with a variety of job types and descriptions at levels from post-baccalaureate to PhD[3]. Employment roles are also differentiated by a continuum of expertise in topics of astronomy or computing. This range of data support positions requires a diversity of training opportunities in order to work at the various levels. In addition, different career development and advancement suited to those career tracks is also required. For example, positions for individuals with PhDs are significantly different from those positions that require only a post-bac degree, and thus the metrics used to support and determine career advancement must also be different. It has only recently been recognized that these roles should be trained for and tracked independently of scientific interests and other professional duties. Consequently, the community has not adequately tracked the quantity and demographics of astronomy researchers currently engaged in science data support roles. Instead of quoting statistics here, we present exemplar descriptions of current job titles and duties.

Many of the people engaged in science data support hold PhDs in astronomy, astrophysics or physics. These researchers may be employed at colleges, universities, data centers, observatories, or national laboratories. They may hold a leveled academic titles (e.g., Professor, Astronomer, Scientist, etc.), as well as an additional functional job position in centers or programs with names "Data Science Mission Office", "Community Science and Data Center", "Infrared Processing and Analysis Center." Even though the scientists in support roles, have distinct duties and responsibilities, the same traditional metrics that quantify the number and impact of publications are used to evaluate their productivity. Clearly, we need to also create new measures that appropriately capture the output of these new, non-traditional roles.[4]

---

[3] For instance, according to the AAS's 2016 membership survey, 18% of non-postdoctoral members and 47% of postdocs employed at universities and 4-year colleges engage in simulations, data mining, data visualization, software and/or data analysis as the main activities of their work.

There are also many other science data support roles, in which staff have degrees at the BS, MS, or PhD level with position titles like "research and instrument associate", "research and instrument scientist", "mission systems scientist", or "archive scientist." Such staff members are often responsible for coding and database support. Below, we discuss the resources and cultural changes needed to support the career trajectories of this workforce, to slow the threat of "brain drain" from the field, and to develop a workforce that can thrive in academia, industry, or government lab positions.

**4. Training to contribute to software development: Building the next generation**

Astronomers have a long history of developing useful software, but software development itself has not been considered a core competency of the astronomy training curriculum. The expectation of petascale datasets in the 2020's provides a strong motivation to increase familiarity with effective practices in software development, as well as the adoption of software frameworks that are widely used in the commercial sector. Enhanced software skills not only improve the quality of astronomical software, it can also accelerate both the scale and pace of scientific discovery.[5] In addition, such software training creates a more skilled workforce who have increased familiarity with (and marketability to) opportunities in industry.

Currently, industry standard practices include using version control (e.g., GitHub), maintaining documentation and unit tests with code, and employing continuous integration methodologies, in which code is built and executed in shared repositories, allowing teams to identify issues early. However, the typical astronomer is likely not employing these practices. Astronomical projects are now comparable in scale to large industrial software development projects. Analysis in the 2020s will involve many pieces of software that are integrated into complex pipelines, processing ever-larger volumes of data. Consequently, the gap between these effective practices and the modern cultural norm in astronomy must be reduced as the field transitions to increasingly large collaborations.

---

[4] See Astro2020 APC white paper,"Elevating the Role of Software as a Product of the Research Enterprise", Arfon Smith, et al. for additional details.
[5] For examples, see Astro2020 Science White Papers listed in the References section.

The increasingly critical role of software development in astronomy makes it crucial to include software development as a part of the core graduate curriculum alongside typical coursework, like mathematics and observing techniques[6]. Such coursework will also help reduce the disparity between students from different backgrounds, some of whom may never have been exposed to software development, or even coding, as undergraduates. This course material complements, but is distinct from, training in data science and scientific computing techniques, which are increasingly being incorporated into astronomy coursework. Developing the course material for data science work is likely beyond the scope of most departments, but vital steps have already been taken by several groups. Notably, the LSST Data Science Fellowship Program has already developed materials to expose students to best practices for software development.[7] Curating these materials, and augmenting them with information for distribution on widely-used platforms will reduce the barrier to adopting such coursework or integrating it into existing classes. Other grassroots training opportunities also exist, but in a similarly piece-meal and uncoordinated way that fails to address the long-term requirements of the field into the next decade.
Recommendations 1.1.1, 1.2.3, 1.2.4 and 1.2.5 address these issues.

Yet another challenge resides in the structure and processes of university departments. Many computer science departments do not have classes set up for scientists who are non-majors or do not teach the programming skills necessary for scientists. Thus, the burden of developing more appropriate materials is fractured and currently falls upon individual instructors. Current faculty, especially at smaller, under-resourced schools, do not have the resources to develop these more focused courses. The field needs dedicated staffing to develop curriculum materials for computational training. A fundamental barrier to the development of reliable, curated, and widely shared software in astronomy is the lack of incentives for this work and the dominance of the "publish or perish" mentality. Changing this cultural norm requires that our community incentivize --- both within scientific projects and across the field at the employment level --- work in developing good software and in educating people to build good software. Two key updates must be made to our community's value structure to change cultural norms and prepare for the software challenges of projects in the 2020's: 1) software development work should be a part of assessments in service and research; 2) software that is widely used for scientific work should be well-written and documented. A full solution cannot be realized through universities alone, and partnerships with data centers, observatories, national labs, and professional societies are crucial. Funding agencies have a key role to play in leading this evolution. Recommendations 1.1.2,1.2.3, 1.2.4 and 1.2.5 address these concerns.

---

[6] For additional details see ASTRO2020 APC white paper, ``Training the Future Generation of Computational Researchers'', Gurtina Besla, PI
[7] See https://astrodatascience.org/curriculum

The popularity of the existing training programs, which grew organically out of the community, attest to the need for additional and more advanced training resources. While there are several successful programs that address some of these concerns, they are insufficient to meet the needs of the larger community.  For example, the Software Carpentry curriculum[8] is limited to the very basics of version control and collaborative software development but does not cover topics, like performance optimization, continuous integration and testing, or documentation. Furthermore, most of these workshops are targeted to senior graduate students, with a few targeting early-career scientists, and they are not designed to meet the needs or concerns of mid-career scientists and managers. Thus, these programs are currently limited to a very small portion of the community and are currently unable to  provide the needed training to people in multiple sectors of our community who need and want these opportunities. Recommendations 1.1.1, 1.1.2 and 1.3.6 suggest solutions for these concerns.

Besides developing course material, there are several other challenges to supporting scientific  software  training  in a university setting. Another hurdle is the lack of access to state-of-the-art technologies. Because the landscape of coding and software development changes rapidly as coding languages come and go, workflow best practices continually evolve, and new platforms emerge and gain wide acceptance. For principal investigators and project managers to make informed decisions and guide their teams, there must be opportunities to stay abreast of these developments and to evaluate their utility even if these managers are not the ones actually using these tools. Recommendation 1.2.4 addresses this need.

Staff at data centers may themselves currently lack up-to-date data science skills and knowledge. Funding to support career development for current staff and to provide resources for centers to hire staff that have data science expertise is critical to building workforce capacity in the 2020s. (See Recommendations 1.2.4, 1.3.6)

**5. Training to take advantage of big data for research: Bytes to Science**

---

[8] https://software-carpentry.org/lessons/

Astronomers who came of age before the era of Big Data require training to take advantage of astronomical data in the 2020s. They also need these skills to mentor students who are simultaneously learning both astrophysics and the uses of data for research. It is crucial that access to training be made widely available to these professionals who come from a variety of science backgrounds and are based at a broad range of institutions (e.g., universities, data centers, etc.). This is especially important, considering these professionals will be cultivating the next generation of scientists, as well as making decisions about investment in new technologies. If access to advancing data skills remains difficult to obtain, we will fail to build a diverse workforce equipped to answer the most pressing questions in astronomical research.[9] With proper support, data centers could play an important role in providing this training. Recommendations 1.3.6 and 1.1.2 highlight this role.

New, freely accessible open source code and Jupyter frameworks, like SciServer.org and NOAO Data Lab, enable anyone with a web browser to quickly and easily analyze vast stores of professional astronomy datasets via web-based notebooks. These cloud-based platforms can democratize educational access by providing a scale of computing power and data storage that was previously reserved for students and faculty at well-resourced research institutions, where high-performance computing access and support are abundant.[10] A small number of astronomers in higher education are already developing instructional activities for these platforms. These materials train students and other users to explore and analyze large professional astronomy datasets with ease and to equip users with the computational foundation needed to pursue advanced independent research projects. (Recommendations 1.1.2, 1.2.3,1.2.4 and 1.3.7 address these concerns.)

Jupyter notebooks in particular hold enormous potential for training the current and next generation of astronomy professionals. However, currently, standardized curricular activities are developed in an entirely ad-hoc manner with no reporting or assessment of their effectiveness. Limited resources (funding and time) lead to very little deliberate coordination amongst various astronomy faculty who produce such materials, and these products are not sufficiently discoverable (e.g. accessible through a common repository). The establishment of community science centers hosted by data centers (like NOAO) can be a hub (clearing house) to bring information to the community about opportunities for the kind of resources and training that allow a broad group of researchers to go from petabytes to publications.
(See Recommendations 1.3.6 and 1.2.4.)

---

[9] Several ASTRO2020 science white papers directly discuss the software needs of the next decade. We list some of these papers in the references section.
[10] Norman, Dara J., "Can Big Data Lead an Inclusion Revolution?", ASTRO BEAT: Inside the ASP, 162, July 2018.

In order to provide the most useful training, data centers need a clear view of user needs. This information is provided by advisory committees, like "User Panels." However, these panels are traditionally populated by astronomers based at R1 institutions and other data centers. Data centers should ensure that their user panels include representatives from small and under-resourced institutions; this will provide a clearer picture of the unique training needs and challenges that must be addressed for these researchers. In addition, community surveys of astronomers who do not currently use data centers, should be undertaken to better understand what barriers exist. Recommendation 1.3.7 addresses this concern.

**6. Astro2020 Science Goals Require a Tech-Savvy Workforce**

In the submitted Astro2020 **science** white papers, more than 30 directly mention software as a crucial part of achieving science goals in the next decade. An even greater number allude to software packages, pipelines and methods that will need to be developed or matured to support the science goals of research on exoplanets, stellar populations, galaxy evolution, multi-messenger astrophysics and more topics in the 2020s. It is essential that we have a workforce that is, prepared, equipped and maintained to build, support and use this technical infrastructure to reach these science goals. It is also urgent that we draw this needed workforce from a diverse pool of expertise and abilities in order to achieve the best science[11].

**7. References: ASTRO2020 Science White Papers**

1. Behroozi, Peter., et al., Empirically Constraining Galaxy Evolution'
2. Chang, Philip, et. al., 'Cyberinfrastructure Requirements to EnhanceMulti-messenger Astrophysics'
3. Chary,Ranga-Ram, et al., Cosmology in the 2020s Needs Precision and Accuracy: The Case for Euclid/LSST/WFIRST Joint Survey Processing'

---

[11] e.g., Page, Scott,"The Diversity Bonus", Princeton and Oxford: Princeton University Press, 2017.

4. Fabbiano, et al.,'Increasing the Discovery Space in Astrophysics: The Exploration Question for Resolved Stellar Populations'
5. Fabbiano, et al.,'Increasing the Discovery Space in Astrophysics: The Exploration Question for Stars and Stellar Evolution'
6. Fabbiano, et al.,'Increasing the Discovery Space in Astrophysics:;The Exploration Question for Planetary Systems'
7. Fabbiano, et al.,'Increasing the Discovery Space in Astrophysics:The Exploration Question for Compact Objects'
8. Fryer, Chris L., et al., 'Catching Element Formation In The Act The Case for a New MeV Gamma-Ray Mission: Radionuclide Astronomy in the 2020s
9. Greene, Jenny, et al.,'The Local Relics of Supermassive Black Hole Seeds
10. Law Casey J., et al., 'Radio Time-Domain Signatures of Magnetar Birth
11. Olsen, Knut, et al., 'Science Platforms for Resolved Stellar Populations in the Next Decade'
12. Rhodes, Jason, et al., 'Cosmological Synergies Enabled by Joint Analysis of Multi-probe data from WFIRST, Euclid, and LSST
13. Siemiginowska, Aneta, et al., 'The Next Decade of Astroinformatics and Astrostatistics''

10